\documentstyle[floats,aps,epsf,prl]{revtex}

\def \be{\begin{displaymath}}
\def \ee{\end{displaymath}}              

\def \ben{\begin{equation} }
\def \een{\end{equation}   }            

\def \bea{\begin{eqnarray*}}             
\def \eea{\end{eqnarray*}}

\def \bean{\begin{eqnarray}}             
\def \eean{\end{eqnarray}}

\def \Ref#1{(\ref{#1})}
\def \eps{\varepsilon}

\def \fie{\varphi}
\def \e{ {\rm e}}

\def \inv{ ^{-1} }

\def \invb#1 { \frac{1}{#1} }

\def \av#1{ {\left\langle #1 \right\rangle} }
\def \ave#1{ {\left\langle #1 \right\rangle}_{{\rm\scriptsize e}}}
\def \avr#1{ {\left\langle #1 \right\rangle}_{{\rm\scriptsize res},m}}

\def \fr#1#2{ \frac{#1}{#2} }
\def \binom#1#2{ \left( {#1} \atop {#2} \right) }

\begin{document}
\draft 
\twocolumn[\hsize\textwidth\columnwidth\hsize\csname @twocolumnfalse\endcsname

\title{Point-Contact Conductances from Density Correlations}

\author{Rochus Klesse and Martin R. Zirnbauer}

\address{Universit\"at zu K\"oln, Institut f\"ur Theoretische Physik,
  Z\"ulpicher Str.~77, D-50937 K\"oln, Germany.}
\date{Sept 30, 2000}

\maketitle

\begin{abstract}
  We formulate and prove an exact relation which expresses the moments
  of the two-point conductance for an open disordered electron system
  in terms of certain density correlators of the corresponding closed
  system.  As an application of the relation, we demonstrate that the
  typical two-point conductance for the Chalker-Coddington model at
  criticality transforms like a two-point function in conformal field
  theory.
\end{abstract}
]

In mesoscopic physics, as well as in quantum physics at large, {\it
  open} systems are distinguished from {\it closed} ones.  The
difference is most evident from the nature of the corresponding energy
spectra: when a closed and finite system is opened up, the discrete
set of stationary solutions of the Schr\"odinger equation turns into a
continuum.  In the open case, one usually looks for solutions in the
form of scattering states, where the wave amplitudes are prescribed in
all incoming channels and the outgoing waves are then related to the
incoming ones by a linear operator called the scattering matrix or
$S$-matrix.

One may ask whether one can connect the scattering matrix of an open
system with the properties of its closed analog.  In spite of the
physical distinction between open and closed systems, such connections
do exist.  For example, for noninteracting quasi-one-dimensional
electrons subject to weak disorder it has been found \cite{mjp} that
the parametric correlations between the eigenphases of the $S$-matrix
exhibit behavior similar to that of the energy levels of the closed
system.  For another example, the $S$-matrix (or rather its extension
including evanescent modes) has been utilized \cite{ds} in numerical
algorithms that compute large numbers of energy eigenvalues for a
chaotic closed billiard.  An intimately related scheme is provided by
Bogomolny's semiclassical T-operator \cite{bogomolny}.

All these theoretical developments were primarily concerned with the
energy levels and their statistics.  In the present Letter we address
a different question: suppose a closed system has been opened, 
what relations, if any, exist between the statistical properties
of the $S$-matrix and the stationary {\it wavefunctions}
of the closed system? This question is of practical
relevance, as the $S$-matrix determines the electrical conductance of
an open mesoscopic device. It will be seen that, provided a few
well-formulated conditions are satisfied, an informative answer can be
given. 
While this seems unlikely if the system is perturbed strongly,
say by making the boundaries transparent, the situation is different
if one uses point contacts which establish the connection to  
the outside through ideal one-dimensional channels. For concreteness
and simplicity, we are going to develop the theory for the case where
time is a discrete variable, but we anticipate that a variant of the
relation to be derived remains valid for systems with continuous-time
dynamics generated by a Hamiltonian. 

We begin by briefly reviewing the concept of network models.  Wave or
quantum-particle propagation in a random medium or a chaotic cavity
can be modeled by a network (or graph) consisting of internal channels
represented by links and of local scattering centers represented by
nodes.  For a network with $N$ links, the state is specified by $N$
complex amplitudes that assemble into a vector $\Psi = \{ \psi_l \}_{l
  = 1, \cdots , N}$.  The network evolves in time by repeated
application of a unitary matrix $U \in {\rm U}(N)$, which propagates
states between integer times $t$: $\Psi(t+1) = U \Psi(t)$.  Stationary
states $\Psi_j = \{ \psi_{j ,l} \}$ and quasi-energy levels $0 \le
\omega_1 \le \dots \le \omega_N < 2\pi $ are constructed by solving
the eigenvalue problem $U\Psi_j = {\rm e}^{i\omega_j} \Psi_j$.  By the
local intensity $\rho_{j,l}$ of $\Psi_j$ we mean the squared amplitude
$|\psi_{j,l}|^2$ of the normalized state $\Psi_j$ at link $l$.

\begin{figure}
  \begin{center}
    \epsfxsize=8cm \leavevmode \epsffile{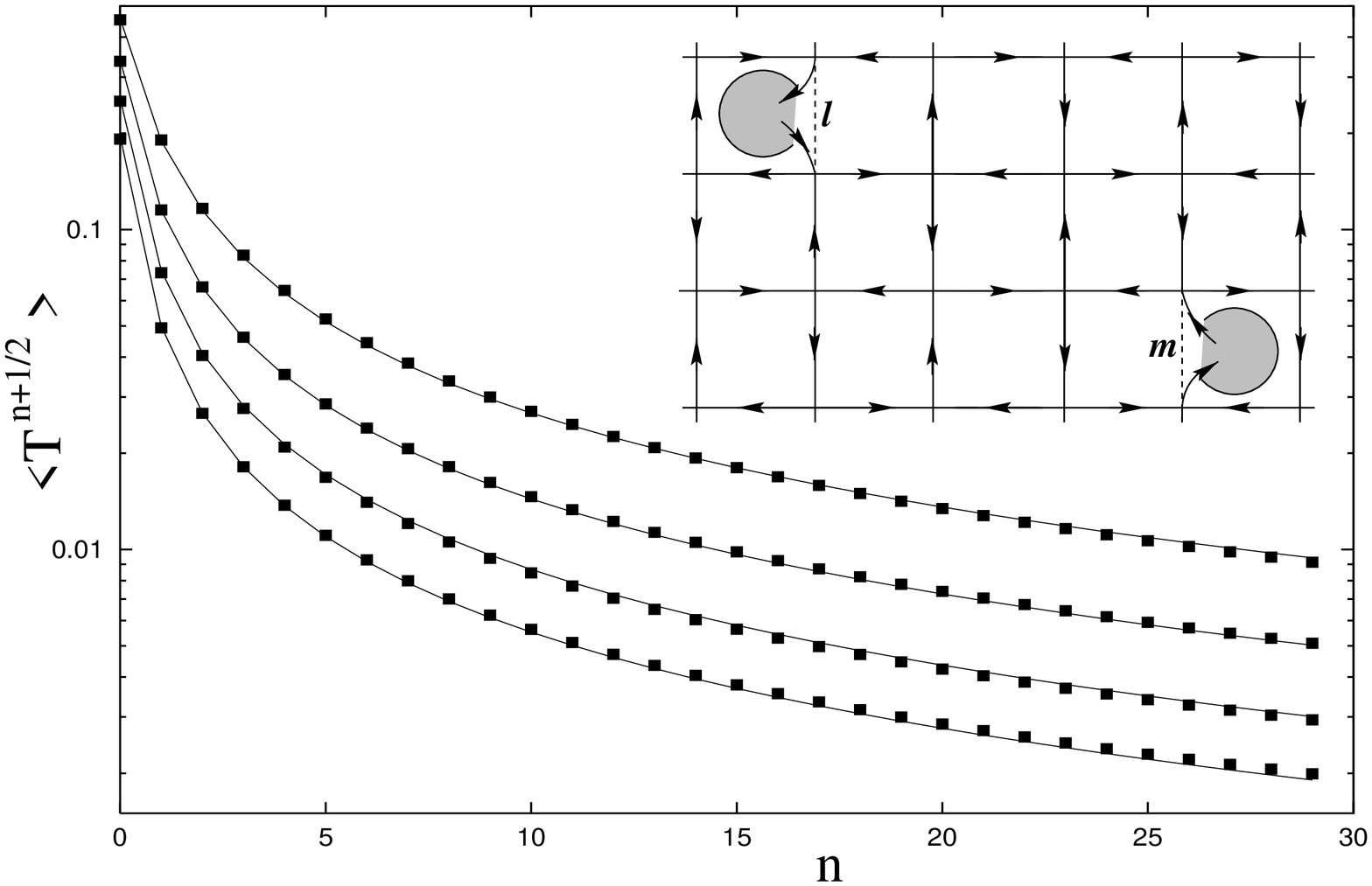}
    \caption{Half-integer moments $\av{T^{n+1/2}}$ for $n = 0, \dots,
      29$ of the two-point conductance of the Chalker-Coddington model
      at criticality.  The four data sets correspond to contact
      distances $x = 5, 10, 15$ and 20 (top to bottom) in a cylinder
      geometry with length $L = 100$ and circumference $W = 10$.
      Solid lines represent the theoretical prediction.  Inset:
      Chalker-Coddington network connected to external charge
      reservoirs via two interior point contacts. }
    \label{fig1}
  \end{center}
  \vspace{-0.5cm}
\end{figure}

The previous definitions apply for closed networks, where each link
starts and ends at a network node.  To define a conductance, however,
one needs to connect the network to external charge reservoirs.  We do
this by introducing two interior {\it point contacts}, i.e.~by cutting
the network open at two links $l$ and $m$, thereby producing two
incoming and outgoing channels. If we imagine these channels to be
connected to Ohmic contacts, the upshot of the surgery done on the
network is to define a two-terminal conductance between the points $l$
and $m$.  We denote it by $T_{lm}$ and refer to it as the two-point
conductance for short.  The inset of Fig.~\ref{fig1} shows the basic
setup at the example of the Chalker-Coddington network \cite{CC}.

Given the $2\times 2$ scattering matrix $S = {\scriptsize \pmatrix{
    S_{\tiny ll} &S_{\tiny lm}\cr S_{\tiny ml} &S_{\tiny mm}\cr}}$
which consists of the amplitudes for transmission and reflection by
the open network, the Landauer-B\"uttiker formula states that $T_{lm}$
(at zero temperature, and in units of the conductance quantum $e^2/h$)
equals the transmission probability $|S_{lm}|^2$.  For the following
it will be crucial that $S$, though primarily defined for the open
network, can also be expressed \cite{jmz} entirely in terms of the
unitary time-evolution operator $U$ of the {\it closed} system:
\begin{equation}
  S_{\alpha\beta} = \left\langle \alpha \left| (1 - U P_l P_m)^{-1} U
    \right| \beta \right\rangle \quad (\alpha,\beta=l,m)
  \label{smatrix}
\end{equation}
where $P_k = 1 -|k\rangle\langle k|$.  Expansion of $(1 - U P_l P_m)
^{-1}$ in a geometric series shows that $S_{\alpha\beta}$ is a sum
over path amplitudes for paths from $\beta$ to $\alpha$.  The
projection operators $P_l$ and $P_m$ serve to truncate paths as soon
as they arrive at the outgoing channel of either terminal $l$ or $m$.

Via relation (\ref{smatrix}), every disorder average $\av{\cdots}$ of
the closed system determines a disorder average pertaining to the
$S$-matrix of the open system.  In particular, the average of any
function of the two-point conductance is defined, by $\av{F(T_{lm})}
\equiv \av{F \left( \big| \langle l|(1-U P_l P_m)^{-1} U |m \rangle
    \big|^2 \right)}$.

Starting from some disordered network model, our aim in the sequel is
to relate the distribution of two-point conductances to the statistics
of stationary wavefunctions.  As will be seen, the relation we are
going to derive does not depend on the details of the model chosen.
The only requirements we need are two conditions constraining the
disorder average: the phases of the reflection coefficients $S_{ll}$
and $S_{mm}$ must be (i) {\it uniformly distributed and independent of
  each other}, and (ii) {\it statistically independent of $T_{lm}$}.
These conditions are automatically satisfied for the
Chalker-Coddington model and more generally for all models with local
phase invariance.

Given some choice of model with disorder average $\av{\cdots}$, we
introduce an additional average $\ave{\cdots}$ for observables that can
be expressed by the stationary states $\Psi_j$:
\begin{equation}
  \label{ES}
  \ave{A} := \frac{1}{\eps\nu} \int_{0}^{\eps} d\omega \sum_{j=1}^N
  \left\langle \delta(\omega-\omega_j) A(\Psi_j) \right\rangle \;,
\end{equation}
where $\nu$ is the level density at $\omega = 0$, and $\eps$ sets the
size of an energy window.  In what follows, we take $\eps \to 0$,
restricting the sum over states to the first level, $\omega_1$.

We are now in a position to propose a statistical relation between
$T_{lm}$ and the intensities $\rho_l = |\psi_{1,l}|^2$ and $\rho_m =
|\psi_{1,m}|^2$.  Let $f(x)$ be any function defined for $x$ in the
positive real numbers.  Then we claim \cite{symmetry}
\begin{eqnarray}
  &&2\pi\nu \: \ave{\rho_m f\left( {\rho_m}/{\rho_l} \right) } =
  \av{F(T_{lm})}\:, \quad {\rm where} \label{relation} \\ &&F(T) :=
  \int_0^{2\pi} \frac{d\varphi}{2\pi} \: f\left( T^{-1} |1 - {\rm
      e}^{i\varphi}\sqrt{1-T}|^2 \right)\:. \label{trafo}
\end{eqnarray}
A first remark on \Ref{relation} is that, in view of the scaling $\nu
\sim N$ and $\rho_m \sim N^{-1}$ (on average over the statistical
ensemble), the system size $L^d \sim N$ drops out of the combination
$\nu\rho_m f(\rho_m/\rho_l)$, reflecting the fact \cite{mrz94} that
$T_{lm}$ tends to a finite limit as $L$ is sent to infinity.  Note
that the plain correlator $\nu^2 \av{\rho_l\rho_m}$ does not share
\cite{cd} this property of size independence, which is why any attempt
to express it through such relations as (\ref{relation}-\ref{trafo})
is met with failure. 

More explicit relations are obtained by adopting special choices for
$f$.  Of particular interest is an expression giving the typical
conductance $T_{lm}^{\rm typ} \equiv \exp\av{\ln T_{lm}}$:
\begin{equation} 
  \label{lnT}
  \av{\ln T_{lm}} = 2\pi\nu \: \ave{\rho_m \ln(\rho_l / \rho_m)} \:,
\end{equation}
which follows from (\ref{relation}-\ref{trafo}) on taking $f(x) = -\ln
x$. Another simple relation results on setting $f(x) = (1+x)^{-1}$:
\begin{displaymath}
  \langle \sqrt{T_{lm}} \rangle = 4\pi\nu \: \ave{\rho_l \rho_m /
    (\rho_l + \rho_m)} \:.
\end{displaymath}
Furthermore, by inverting the relation gotten by setting $f(x) = x^n$,
and analytically continuing from integer $n$ to complex $z$, it can be
shown with some effort that
\begin{displaymath}
  \av{T_{lm}^z} = \sum_{k=0}^\infty \frac{(2k+1)\: \Gamma(1-z)^2 \:
    2\pi\nu}{\Gamma(k+2-z)\Gamma(1-k-z)} \ave{\rho_m^{k+1} \rho_l^{-k}
    } \:.
\end{displaymath}
Note that the sum over $k$ is terminated by the poles of the gamma
function when $z$ is a negative integer or zero.  For all other values
of $z$, the series will in general be divergent, since the polynomial
decay of the product of inverse gamma functions does not suffice to
control the growth with $k$ of the moments $\langle \rho_m^{k+1}
\rho_l^{-k} \rangle$.  Fortunately, if $z = n+1/2$ with $n$ a positive
integer, the series can be rearranged to produce a finite sum
involving polynomials in the bounded random variable $\rho_l \rho_m /
(\rho_l + \rho_m)^2 \le 1/4$:
\begin{equation}
  \langle T_{lm}^{n+1/2} \rangle = {\binom{2n}{n}}^{-1}
  \sum_{k=0}^{n} \binom{2k}{k} {\mu_{n-k} \over 1-2k} \;,
  \label{moments}
\end{equation}
where $\mu_n = 2^{4n+2} \pi\nu \ave{(\rho_l \rho_m)^{n+1} (\rho_l +
  \rho_m)^{-2n-1}}$.  We will put this formula to good use below.

Our main result is relation \Ref{relation}.  Its proof is based on the
observation that for a network at resonance, i.e.~for $\omega_1 = 0$,
the corresponding eigenvector $\Psi_1$ of $U$ can be interpreted as a
scattering state of the network opened at the links $l$ and $m$.  In
other words, the two amplitudes $\psi_{1,l}$ and $\psi_{1,m}$ of an
eigenstate $\Psi_1 = U \Psi_1$ constitute at the same time an
eigenvector $(\psi_{1,l}, \psi_{1,m})$ with eigenvalue unity of the
scattering matrix $S$.  Thus we have $S_{ll} \psi_{1,l} + S_{lm}
\psi_{1,m} = \psi_{1,l}$ and therefore, assuming $\psi_{1,l} S_{lm}
\neq 0$,
\begin{equation}
  \frac{\rho_m}{\rho_l} = \frac{|1-S_{ll}|^2}{|S_{lm}|^2} = \frac{|1 -
    \e^{i\arg{S_{ll}}} \sqrt{1-T_{lm}}|^2}{T_{lm}} \;. \label{ratio}
\end{equation}

The next step will be to take the disorder average of this equation,
viewing both sides as functions on the closed network.  Of course,
since \Ref{ratio} holds only for networks at resonance, it is
imperative that we restrict the disorder average to the resonant
realizations.  We do this by tuning the random phase $\varphi_m$ at
link $m$ to a function $\varphi_m = \varphi_m^0(\xi)$ of all other
random variables $\xi = \{ \xi_1 , \xi_2 , ... \}$ so that $\omega_1
(\xi, \varphi_m^0 (\xi)) = 0$ is identically satisfied for all $\xi$
\cite{remark}. The restricted or resonant disorder average thus
defined is denoted by $\avr{\cdots}$. By applying some function $f$ to
the identity \Ref{ratio} and averaging in this manner we get
\begin{displaymath}
  \left\langle f \left( \frac{\rho_m}{\rho_l} \right)
  \right\rangle_{{\rm res},m} = \avr{ f\left( \fr{|1 - \e^{i\varphi_l}
        \sqrt{1 - T_{lm}} |^2}{T_{lm}} \right) } ,
\end{displaymath}
where we have set $\varphi_l \equiv \arg{S_{ll}}$.  Now recall that,
according to conditions (i) and (ii), the phase $\varphi_l$ is
statistically independent of $\varphi_m$ and $T_{lm}$ and is uniformly
distributed.  Consequently, the preceding equation continues to hold
if we integrate the r.h.s.~with measure $d\varphi_l / 2\pi$ over
$[0,2\pi]$, thereby turning it into $\avr{F(T_{lm})}$ with $F$ defined
by \Ref{trafo}.  Moreover, condition (ii) entails that the resonant
average $\avr{F(T_{lm})}$ actually equals the natural average
$\av{F(T_{lm})}$.  Thus
\begin{equation}
  \avr{f(\rho_m/\rho_l)} = \av{F(T_{lm})} \:. \label{res_av}
\end{equation}

Next we transform the average over resonant networks on the l.h.s.~to
the eigenfunction average \Ref{ES}, as follows:
\begin{eqnarray*}
  &&2\pi\nu \: \ave{A} = \lim_{\eps\to 0} \frac{2 \pi} {\eps}
  \int_0^\eps d\omega \: \av{\delta(\omega_1-\omega) A(\Psi_1)} = \\ 
  &&\int\limits_0^{2\pi} d\varphi_m \: \av{\delta \big(\omega_1(\xi,
    \varphi_m)\big) A(\Psi_1)}_{\xi} = \avr{{A(\Psi_1) \over
      |\partial\omega_1 / \partial\varphi_m|}} .
\end{eqnarray*}
Setting $A = \left|{\partial\omega_1 \over \partial\varphi_m} \right|
f(\rho_m / \rho_l)$ and using \Ref{res_av} we arrive at
\begin{displaymath}
  \av{F(T_{lm})} = 2\pi\nu \: \ave{ \left|\frac{ \partial\omega_1}
      {\partial \varphi_m}\right| f{\scriptsize \left( \frac{\rho_m}
        {\rho_l}\right) }} \:.
\end{displaymath}
Finally, since a phase variation $\delta\varphi_m$ perturbs the
time-evolution operator $U$ by $i(\delta\varphi_m) |m\rangle\langle m|
U$, standard first-order perturbation theory shows that the level
velocity $|\partial\omega_1 / \partial\varphi_m|$ equals $\rho_m$.
This concludes the proof.

It is easily seen that the above line of reasoning does not depend in
any essential way on assuming a network model with discrete time.  In
fact, a slight variant of the argument goes through for
continuous-time dynamics generated by a Hamiltonian $H$.  What one
needs to do is attach two identical and strictly one-dimensional
leads $l$ and $m$ 
to the system, which are then either connected to charge reservoirs,
or else closed off by imposing reflecting boundary conditions at
variable lengths $L_l$ and $L_m$.  The intensity $\rho_{l(m)}$ of the
closed system is interpreted as $\int_{x_0}^{x_0+\lambda} dx\;|\psi(x)|^2
$ where $\lambda$ is the wavelength and $x_0,x_0+\lambda$ are in the
closed lead $l(m)$. The normalization factor $2\pi \nu$ in
Eq.~\Ref{relation} has to be replaced by $\av{\rho_m}\inv$.
To tune the system to a given resonance energy $E$, we adjust $L_m$.

At the moment it is not clear to us to which extent, if at all,
the relation applies to systems with more than two point contacts or
more general types of contact. This point deserves further
investigation. 

When states are localized, relation \Ref{lnT} implies that the decay
length of the typical two-point conductance agrees with the typical
localization length of the states of the closed system.  While this
has long been known for strictly one-dimensional systems
\cite{borland}, it is now seen to hold exactly under the much more
general conditions we have specified.  By word of caution, however, we
recall that our formula does {\it not} cover correlators like $\nu^2
\langle \rho_l \rho_m \rangle$, so it does not settle a recent debate
as to whether two-scale localization in a magnetic field can occur
\cite{ke} for closed systems when it does not \cite{sb} for open ones.

We now demonstrate the utility of the result (\ref{relation}) by
applying it to a challenging problem: critical transport at the
integer quantum Hall (IQH) transition \cite{huckestein}.  To make
analytical progress with that transition, one needs to confront the
question what can be learned from conformal field theory.  For
critical systems without disorder, the principle of conformal
invariance has proven to be very powerful, by posing strong
constraints on correlation functions and determining how they
transform between various geometries \cite{cardy}.  The question
whether the same principle applies to 2d disordered critical points
was first raised by Chalker \cite{chalker}.  The answer turned out to
be negative for the correlation functions he considered.  However, as
we now proceed to demonstrate, conformal invariance does govern the
two-point conductances.

An analytical formula for the distribution of two-point conductances
$T \equiv T_{lm}$ for critical quantum Hall systems in the infinite
plane was proposed in \cite{jmz}.  Given that formula, conformal field
theory predicts the probability measure $d\mu(T)$ for the conductance
between two points with coordinates $(0,0)$ and $(x,y)$ on an
infinitely long cylinder of circumference $W$ to be given by
\begin{eqnarray}
  d\mu(T) &=& p(T\inv - 1) T^{-2} dT \;, \label{measure} \\ p(\rho)
  &=& { 2 \pi^{-1/2} \ \zeta^{-X_{\rm t}/4} \over (X_{\rm t} \ln
    \zeta)^{3/2} } \int\limits_{{\rm arsinh} \sqrt{\rho}}^\infty {{\rm
      e}^{-t^2 / (X_{\rm t} \ln\zeta)} t dt \over \sqrt{\sinh^2 t -
      \rho}} \;, \label{integral} \\ \zeta &=& (W/\pi a) \left| \sinh
    \big( \pi (x + iy)/W \big) \right| \:, \nonumber
\end{eqnarray}
where $X_{\rm t}$ is a critical exponent (unknown as yet from theory),
and $a$ is a nonuniversal microscopic scale that sets the length unit
in which to express the correlation functions of the particular IQH
model considered.

To test this prediction, we numerically calculated the eigenfunctions
of the critical time-evolution operator $U$ of the Chalker-Coddington
model, in a cylindrical geometry with large aspect ratios $L/W$, $100
\le L \le 200$ and $3 \le W \le 30$.  An ensemble of 5000 disorder
realizations was employed.  Using Eq.~(\ref{moments}) we computed the
moments $\langle T^{n+1/2} \rangle$ by averaging over the pairs of
links $(l,m)$ with a given distance $|x|$ and their difference vector
pointing along the axis of the cylinder.  The great advantage of this
indirect approach (in contradistinction with the direct method of
\cite{jmz}) is that a large number $\sim N^2$ of data points can be
extracted from a single eigenfunction $\Psi$.  Since the numerical
effort of calculating $\Psi$ and $T_{lm}$ is comparable, we gain in
efficiency by a factor $N^2 \sim (LW)^2$, which can be many orders of
magnitude.

In Fig.~\ref{fig1} theory and numerics are compared for $W = 10$,
$L=100$, and $|x| = 5, 10, 15, 20$ (in plaquette units).  The
theoretical curves were obtained by numerically integrating
(\ref{measure}) against $T^{n+1/2}$.  We find that an exponent $X_{\rm
  t} = 0.54 \pm 0.01$ (and $a = 0.20$) yields the best fit.  The
statistical errors are estimated to be comparable with the symbol
size.  The agreement is clearly very good, showing that the result
predicted by conformal field theory interpolates correctly between 2d
behavior at short distances and quasi-1d behavior at large scales.

\begin{figure}
  \begin{center}
    \epsfxsize=8cm \leavevmode \epsffile{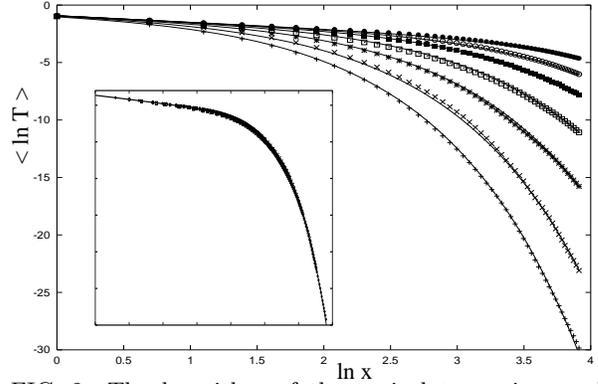}
    \caption{The logarithm of the typical two-point conductance
      $\av{\ln T}$ of the Chalker-Coddington network at criticality
      versus the logarithm of the distance $\ln|x|$, for widths $W =
      3, 4, 6, 9, 14, 20$, and 30 (from bottom to top).  The solid
      lines are functions given by Eq.~\Ref{typical}.  Inset: Rescaled
      curves $\av{\ln T}+ X_{\rm t}\ln W$ versus $\ln |x/W|$ with
      $X_{\rm t} = 0.57 $.}
    \label{fig2}
  \end{center}
  \vspace{-0.5cm}
\end{figure}

The most clear-cut test of conformal invariance is offered by the {\it
  typical} two-point conductance.  By integrating (\ref{measure})
against $\ln T$, one finds a very simple law \cite{jmz}:
\begin{equation}
  T^{\rm typ}_{(x,0);W} = \left| (W/\pi a) \sinh(\pi x/W)
  \right|^{-X_{\rm t}} \:.
  \label{typical}
\end{equation}
Fig.~\ref{fig2} plots $\langle \ln T \rangle$ versus $\ln|x|$ on
cylinders of length $L = 200$ and circumferences $W$ ranging over one
decade, from $3$ to $30$.  We here used 2800 disorder configurations
for each $W$.  In the inset $\langle \ln T \rangle + X_{\rm t} \ln W$
is plotted against $\ln |x/W|$ with an exponent $X_{\rm t} = 0.57$,
determined with an error of $0.05$ by optimizing the match of the
numerical data with the prediction \Ref{typical} ($a = 0.20$).  It is
seen that, within the statistical errors, the data collapse onto a
single curve given by (\ref{typical}).  This means in particular that
the short-distance algebraic behavior $T^{\rm typ} = |x/a|^{- X_{\rm
    t}}$ crosses over to the exponential law $T^{\rm typ} \propto {\rm
  e}^{-\lambda |x|}$ for $|x|\gg W$, with the Lyapunov exponent being
$\lambda = \pi X_{\rm t} / W$. While the relation $\lambda W =$
const.\ has been checked before \cite{janssen}, the present work provides the
first test, for a 2d disordered particle system, of the full
crossover between short- and long-distance behavior predicted
\cite{cardy} by conformal field theory.

The values $0.54 \pm 0.01$ and $0.57 \pm 0.05$ we obtain for the
exponent $X_{\rm t}$ deviate somewhat from the value $X_{\rm t} =
0.61$ found in previous work \cite{jmz}.  We attribute this difference
mainly to the much larger statistical uncertainties in the numerical
analysis of \cite{jmz}.  We also emphasize that the accuracies given
here do not include systematic errors.  Possible sources of these are
the finite values of $W/L$ and $a/W$, which affect the results
obtained from different contact distances in a different way.  We
believe that the deviation between the values obtained from the
moments (0.54) and the typical conductance (0.57) gives a realistic
estimate of these systematic errors.
 
R.K. thanks H.~Moraal for help with the inversion of power
series. This research was supported in part by the DFG,
Sonderforschungsbereich 341.


\begin{thebibliography}{10}
\bibitem{symmetry} Owing to the exchange symmetry $T_{lm} = T_{ml}$
  the l.h.s.~of \Ref{relation} can be symmetrized with respect to $l$
  and $m$, yielding \\ $\pi\nu \ave{ \rho_l f(\rho_l/\rho_m) + \rho_m
    f(\rho_m/\rho_l) } = \av{F(T_{lm})}$.
\bibitem{remark} This can always be done,
  as is shown by the following argument: On the closed network we
  choose an arbitrary closed path containing link $m$. Then we
  vary continuously the scattering parameters at the nodes lying on it
  so that the path becomes a simple loop insulated from the remaining
  network. A straightforward calculation shows that for this
  modified network adjusting $\fie_m$ always suffices to tune one
  level to zero. By reasons of continuity and periodicity, this
  property persists as the nodes on the loop are put back to their
  initial values and the original network is restored.
\end{thebibliography}
\end{document}